\documentclass[superscriptaddress,twocolumn,aps,amssymb,amsmath]{revtex4}

\usepackage{graphicx}
\usepackage{multirow}

\usepackage{color}

\usepackage[T2A,T1]{fontenc}
\usepackage[utf8]{inputenc}
\usepackage[russian,english]{babel}

\usepackage{hyperref}

\definecolor{bluecolor}{rgb}{0,0.,1.}
\definecolor{redcolor}{rgb}{.7,0.,0.}

\begin{document}

\title{Extracting information from S-curves of language change}
\author{Fakhteh Ghanbarnejad}
\thanks{Both authors contributed equally to this work.}
\affiliation{Max Planck Institute for the Physics of Complex Systems, Dresden, Germany}
\email{{fakhteh,gerlach,jmiotto,edugalt}@pks.mpg.de}
\author{Martin Gerlach} 
\thanks{Both authors contributed equally to this work.}
\affiliation{Max Planck Institute for the Physics of Complex Systems, Dresden, Germany}
\author{Jos\'e M. Miotto}
\affiliation{Max Planck Institute for the Physics of Complex Systems, Dresden, Germany}
\author{Eduardo G. Altmann}
\affiliation{Max Planck Institute for the Physics of Complex Systems, Dresden, Germany}

\begin{abstract}
It is well accepted that adoption of innovations are described by S-curves (slow start, accelerating period, and slow end). In this paper, we  analyze how much information on the dynamics of innovation spreading can be obtained from a quantitative description of S-curves. We focus on the adoption of linguistic innovations for which detailed databases of written texts from the last 200 years allow for an unprecedented statistical precision. Combining data analysis with simulations of simple models (e.g., the Bass dynamics on complex networks) we identify signatures of endogenous and exogenous factors in the S-curves of adoption. We propose a measure to quantify the strength of these factors and three different methods to estimate it from S-curves. We obtain cases in which the exogenous factors are dominant (in the adoption of German orthographic reforms and of one irregular verb) and cases in which endogenous factors are dominant (in the adoption of conventions for romanization of Russian names and in the regularization of most studied verbs). These results show that the shape of S-curve is not universal and contains information on the adoption mechanism. (published at "J. R. Soc. Interface, vol. 11, no. 101, (2014) 1044"; DOI: http://dx.doi.org/10.1098/rsif.2014.1044)

\end{abstract}

\maketitle

\section{Introduction}
The term S-curve often amounts to the {\it qualitative} observation that the change starts slowly, accelerates, and ends slowly. Linguists generally accept that {\em ``the progress of language change through a community follows a lawful course, an S-curve from minority to majority to totality.''}~\cite{Weinreich:1968}, see  Ref.~\cite{Blythe:2012} for a recent survey of
examples in different linguistic domains. {\it Quantitative} analysis are rare and extremely limited by the quality of the linguistic data, which in the best cases have {\em``up to a dozen points for a single change''}~\cite{Blythe:2012}. Going beyond qualitative observation is essential to address questions like:

\begin{figure*}
\includegraphics{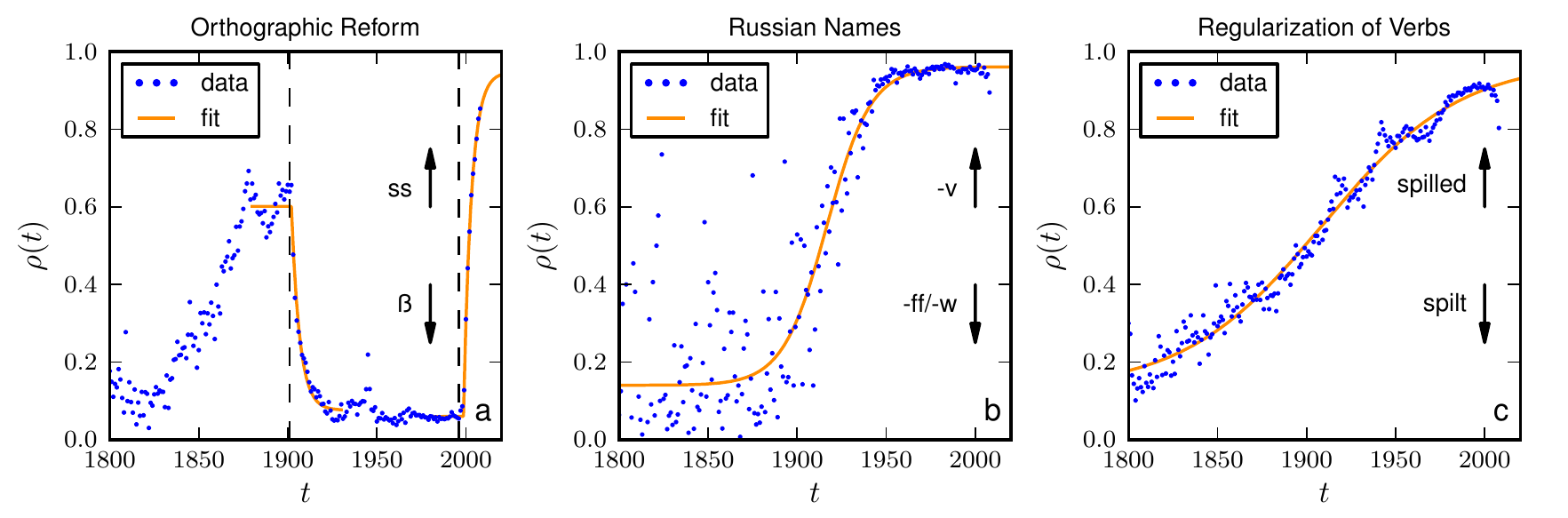}
\caption{(color online) Examples of linguistic changes showing different adoption curves. We estimate the fraction of adopters $\rho(t)$ by the relative frequency as $\rho(t)=\sum_w n^{w}_{1}/\sum_w \sum_q n^{w}_{q}$, where $n^{w}_q$ is the total number of occurrences (tokens) of variant $q$ for the word $w$ at year $t$. (a) The orthography of German words that changed to ``ss'' ($q=1$) from ``\ss'' ($q=2$) in the orthographic reform of 1996 (many words changed from ``ss'' to ``\ss'' in the 1901 reform). (b) The transliteration of Russian names ending with the letter ``\foreignlanguage{russian}{в}'' when written in English (Latin alphabet), changed to an ending in ``v'' ($q=1$) from endings in ``ff''  ($q=2$) or ``w'' ($q=3$) (e.g.,  $w=$ ``\foreignlanguage{russian}{Саратов}'' is nowadays almost unanimously written as ``Saratov'', but it used to be written also as ``Saratoff'' or ``Saratow''). (c) The past form of the verb spill changed to its regular form ``spilled'' ($q=1$) from the irregular form ``spilt''  ($q=2$). The light curve shows the fit of Eq. (\ref{eq:bassSol}). The estimated parameters $a$ and $b$ are  (a) $\hat{a}=0.218,\hat{b}=0.000$ in 1901, and $\hat{a}=0.229$, $\hat{b}=0.000$ in 1996; (b) $\hat{a}=0.000$, $\hat{b}=0.099$; and (c) $\hat{a}=0.001$, $\hat{b}=0.030$. The corpus is the Google-ngram \cite{Michel:2011,Lin:2012} plotted in the minimum (yearly) resolution, see supplementary material (SM) Sec. I for details on the data and Sec. IIIB for details on the fit.
}
\label{fig:extint_lang}\label{fig.1}
\end{figure*}

\begin{itemize}
\item[(i)] Are all changes following S-curves?
\item[(ii)] Are all S-curves the same (e.g., universal after proper re-scaling)?
\item[(iii)] How much information on the process of change can be extracted from S-curves? 
\item[(iv)] Based on S-curves, can we identify signatures of endogenous and exogenous
  factors responsible for the change?
\end{itemize}

Large records of written text available for investigation provide a new opportunity
to quantitatively study these questions in language change \cite{Michel:2011,Lin:2012}.  In Fig. \ref{fig.1} we show
the adoption curves of three linguistic innovations for which words competing for the same
meaning can be identified. Our methodology is not restricted to such simple examples of vocabulary replacement and can be applied to other examples of language change and
S-curves more generally. Here we restrict ourselves to data of aggregated (macroscopic) S-curves because only very
rarely one has access to detailed data at the individual (microscopic) level, see, e.g., Ref.~\cite{Myers:2012} for an exception. 

Data alone is not enough to address the questions listed above, it is also
essential to consider mechanistic models responsible for the change~\cite{Niyogi:2006,Baxter:2006,Ke:2008,Blythe:2012,Pierrehumbert2014}. Dynamical
processes in language can also be described from the more
general perspectives of evolutionary processes~\cite{Blythe:2012,Niyogi:2006,Boyd1985} and complex systems~\cite{Castellano:2009,Baronchelli:2012,Sole:2010}. In this framework, the adoption of new words can be seen as the adoption of innovations~\cite{Rogers:2010,Vitanov:2012,Bass:1969,Bass:2004,Bettencourt:2006,Pierrehumbert2014}. One of the most general and popular models of innovation adoption showing S-curves is the Bass
model~\cite{Bass:1969,Bass:2004}. In its simplest case, it considers a
homogeneous population and prescribes that the fraction of adopters ($\rho$) increases
because those that have not adopted yet ($1-\rho$) meet adopters (at a rate $b$) and are
subject to an external force (at a rate $a$). The adoption is thus described by
\begin{equation} \label{eq:bass}
        \frac{d\rho(t)}{dt} = (a+b \rho(t))(1-\rho(t)).
\end{equation}
The solution (considering $\rho(t_0)=\rho_0$ and $\rho(\infty)=1$)  is
\begin{equation} \label{eq:bassSol}
\rho(t) = \frac{a(1 - \rho_0) - (a+b\rho_0) e^{(a+b)(t-t_0)}  }{-b(1 - \rho_0) - (a+b\rho_0) e^{(a+b)(t-t_0)}}.
\end{equation}
It contains as limiting cases a {\it symmetric} S-curve (for $a=0$) and an exponential relaxation (for $b=0$).
The fitting of Eq.~(\ref{eq:bassSol}) to the data in Fig.~\ref{fig.1} leads to very different $a$ and $b$
in the three different examples, strongly suggesting that the S-curves are not universal
and contain information on the adoption process. For instance, orthographic reforms are known to be exogenously driven (by
language academies) in agreement with $b=0$ obtained from the fit in panel (a).

In this paper we investigate the shape and significance of S-curves in models of adoption
of innovations and in data
of language change. In particular, we estimate the contribution of endogenous and exogenous
factors in S-curves, a popular question which has been addressed in complex systems more generally~\cite{Sornette:2004,Crane:2008,Menezes:2004,Mathiesen:2013}. The different values
of $a$ and $b$ in Eq. (\ref{eq:bass}) are an insufficient quantification, e.g., because
they fail to indicate which factor is stronger.
Here we introduce a definition for the relevance of different factors in a
change. We then show how this quantity can be exactly computed in different models and
propose three different methods to estimate it from the time series of $\rho(t)$.
We compare the accuracy of the methods using simulations of different
network models  and we apply the methods to linguistic changes. We obtain that the exogenous factors are responsible for the change in the German orthographic reforms, but it plays a minor role in the case of romanized Russian names and in most of the studied English verbs which are moving towards regularization.

%
%
%
%
%

\section{Theoretical Framework} \label{sec:GenFram}

Consider that $i=1,\dotsc, N\rightarrow\infty$ identical agents ({\bf assumption 1}) adopt an innovation. The central quantity of
interest for us here is $\rho(t)=N(t)/N$, the fraction of adopters at time
$t$. We assume that $\rho(t)$ is monotonically increasing from
$\rho_0\equiv\rho(0)\approx 0$ to $\rho(\infty)=1$ and agents after adopting the innovation do not
change back to non-adopted status ({\bf assumption 2}).

\subsection{Endogenous and Exogenous Factors}

In theories of language and cultural change, the importance of different factors is
a topic of major relevance, e.g., Labov's internal and external factors \cite{Weinreich:1968} and Boyd and Richerson's different types of
biases in cultural transmission \cite{Boyd1985}. The first question we address is how to measure
the contribution of different factors to the change. To the best of our
knowledge, no general answer to this question has been proposed and
computed in adoption models. As a representative case, we divide factors as endogenous and exogenous to the population. Mass media and decisions from language academies count as exogenous factors while
grassroots spreading as an endogenous factor. In our simplified
classification, Labov's internal (external) factors (to properties of
the language \cite{Weinreich:1968}) are counted by us as exogenous (endogenous), while Boyd and Richerson's \cite{Boyd1985}
direct bias count as exogenous whereas the indirect bias and
frequency-dependent bias count as endogenous.

Our proposal is to quantify the importance of a factor $j$ as the number
of agents that adopted the innovation because of $j$. 
 More formally, let $g_i(t)$ be the adoption probability at time $t$ for agent $i$ (who is
 in the non-adopted status).  We
 assume that $g_i$ can be decomposed in contributions of the different factors $j$ as
 $g_i(t)=\sum_j g_i^j(t)$, where $g_i^j(t)$ is the adoption probability of agent
 $i$ at time $t$ because of factor $j$.   
If $t_i^*$ denotes the time  agent $i$ adopts the
innovation, $g_i^j(t^*_i)/g_i(t^*_i)$  quantifies the contribution of factor $j$ to the
adoption of agent $i$ (the adoption does not explicitly depends on $t<t^*$ and therefore
values of $g_i^j(t)$ for $t<t^*$ are only relevant in the extent that they
influence $g_i^j(t=t^*)$).  In principle, the factor $g_i^j(t^*_i)/g_i(t^*_i)$ can be
obtained empirically by asking recent adopters for their reasons for changing, e.g.,  for
j=exogenous (endogenous) one could ask: {\it How much advertisement  (peer pressure)
  affected your decision?.}
We define the normalized quantification of the change in the
whole population due to factor $j$ as an average over all agents
\begin{equation}\label{eq:def}
G^j=\frac{1}{N}\sum_{i=1}^{N} \frac{ g_i^j(t^*_i)}{g_i(t^*_i)}.
\end{equation}

In order to show the significance of definition~(\ref{eq:def}), and how it can be applied
in practice, we discuss how $g^j_i$ and $G^j$ can be considered in different models. Endogenous (endo) factors happen due to the interaction
of an agent with other agents (internal to the population). They are therefore expected to
become more relevant as the adoption progress (for increasing
$\rho$). Exogenous factors (exo), on the other hand, are related to a source of
information (external to the population) which has no dependence on $\rho$ or time ({\bf assumption 3}). For simplicity, we
report $G \equiv G^{\text{exo}}$ (since $G^{\text{endo}}=1-G^{\text{exo}}$).

\subsection{Population dynamics models}\label{ssec.pop}

Consider as a more general form of Eq. (\ref{eq:bass})
\begin{equation} \label{eq:PopGen}
\dot{\rho}(t) \equiv \dfrac{d \rho(t)}{dt}= g(\rho(t))(1-\rho(t)),
\end{equation}
where $g(\rho(t))$ is the
probability that the population of non-adopters  $(1-\rho(t))$ switches from non-adopted
status (0) to adopted status (1) at a given density of $\rho$. In epidemiology $g(\rho)$ is known as force of infection \cite{Hens:2010}. Since agents are identical (assumption 1) and $\rho(t)$ is invertible (assumption
2), we can associate $g_i^j(t_i^*)$ with $g^j(\rho)$ and $g_i(t_i^*)$ with $g(\rho)$. Introducing
$g(\rho(t))$ from Eq.~(\ref{eq:PopGen}) in the continuous time extension of definition
(\ref{eq:def})  we obtain:
\begin{equation} \label{eq:G}
G^j \equiv \int_0^1 \frac{g^j(\rho)}{g(\rho)} d\rho = \int_{0}^{1}g^j(\rho) \frac{1-\rho}{\dot{\rho}}d\rho = \int_0^{\infty}
 \frac{g^j(t)}{g(t)}\dot{\rho}(t) dt.
\end{equation}
This equation shows that the strength of factor $j$ is obtained by averaging its
normalized strength $g^j(\rho)/g(\rho)$ over the whole population or, equivalently, over
time  (considering the rate of adoption $\dot{\rho}(t)$).

When only exogenous and endogenous factors are taken into consideration, $g(\rho)=g^{exo}+g^{\text{endo}}$ in
Eq. (\ref{eq:PopGen}). Here, assumption 3 mentioned above corresponds to consider that the
adoption happens much faster than the changes in the exogenous factors so that it can be considered independent of time. Therefore $g^{\text{exo}}=g(\rho=0)$. Any change of $g$ with $\rho$ is an endogenous factor and $g^{\text{endo}}(\rho)$ increases with $\rho$ because the pressure for adoption increases with the number of adopters.

For the case of the Bass model defined in Eq.~(\ref{eq:bass}), $g(\rho)=a+b \rho,
g^{endo}=a, g^{exo}=b\rho$ and from Eq.~(\ref{eq:G}) we obtain
\begin{eqnarray} \label{eq:Gextintbass}
G \equiv G^{\text{exo}} = \frac{a}{b} \log_e(\frac{a + b}{a}).
\end{eqnarray}
The correspondence of $a$ and $b\rho$ to exogenous (innovators) and endogenous (imitators) is a
basic ingredient of the Bass model \cite{Bass:1969} \footnote{In our simple model, all agents are identical. The first adopters (innovators) are determined stochastically by the exogenous factor $a$, while agents adopting at the end of the S-curve (imitators) are more susceptible to the endogenous factor $b\rho$.}. However, it is only through Eq.~(\ref{eq:Gextintbass}) that the importance of these factors to the change can be properly quantified. For instance, the case
$a=b$ suggests equal contribution of the factors, but Eq.~(\ref{eq:Gextintbass}) leads to
$G=\log_e 2 \approx 0.69 > 0.5$ and therefore shows that the exogenous factors dominate (are responsible for a larger number of adoptions than the endogenous factors). This new insight on the interpretation of the classical Bass model illustrates the significance of Eq. (\ref{eq:def}) and our general approach to quantify the contribution of factors.

\subsection{Binary state models on networks}\label{ssec.networks}

Another well-studied class of models inside our framework considers agents characterized by
a binary variable $s=\{0,1\}$ connected to each other through a network. We focus on
models with a monotone dynamics (assumption 2), such as the Bass, Voter, and Susceptible Infected
models, which are defined
by the probability $F_{k,m}$ of switching from $0$ to $1$ given that the agent has $k$
neighbours and $m$ neighbours in state $1$~\cite{Newman:2010}. The one dimensional
population dynamics model in Eq.~(\ref{eq:PopGen}) can be retrieved for simple networks
(e.g.,  fully connected or fixed degree). In the general case, we use the
framework of approximate master equations (AME)
\cite{Gleeson:2013,Gleeson:2011} (see SM. II), which describes the stochastic binary dynamics
in a random network with a given degree distribution $P_k$. Assuming as before (assumption 3) that  the
exogenous contribution is given by transitions that occur when no neighbour is infected,
i.e. $g^{\mathrm{\text{exo}}} \left(k,m \right) = F_{k,0}$, we obtain the exogenous
contribution as (see SM. IIB):
 \begin{equation}\label{eq.GAME}
  G = \sum_{k} P_k \sum_{m=0}^k \int_0^{\infty} s_{k,m}  F_{k,0}  \mathrm{d}t,
 \end{equation}
where $s_{k,m}=s_{k,m}(t)$ is the fraction of agents of the $k,m$ class in state $0$.

\begin{figure*}[!bt]
\includegraphics[width=2 \columnwidth]{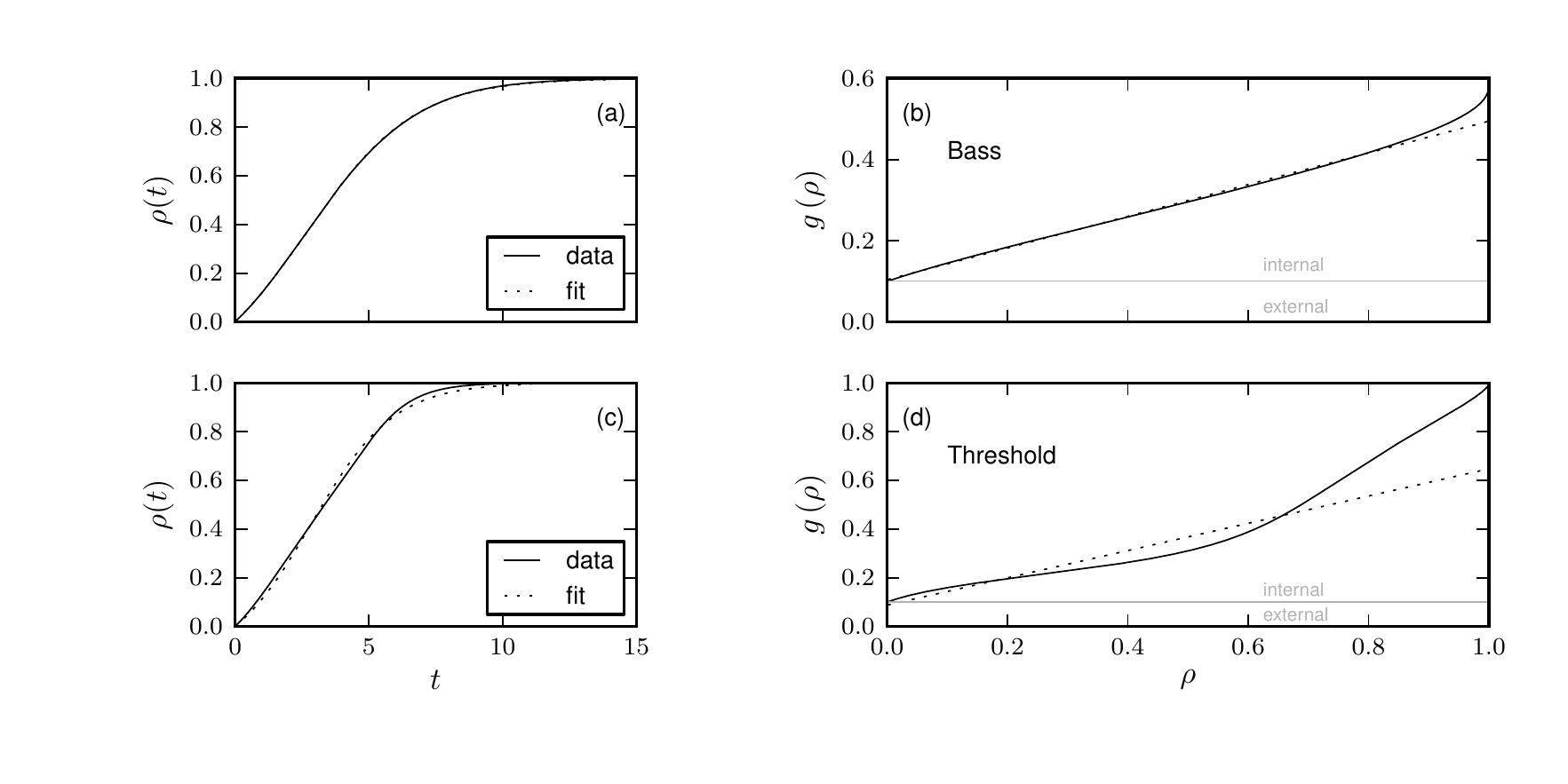}
\caption{Application of time-series estimations to surrogate data. The Bass (a,b) and
  threshold (c,d) dynamics with parameters $a=0.1$ and $b=0.5$ were numerically solved in the AME framework for scale free networks (with degree distribution $P(k)
  \sim k^{-\gamma}$ with $\gamma \approx 2.47$ for $k \in [2,50]$ such that $\langle k \rangle = 4$).  (a,c)
  Adoption curve $\rho(t)$ (fraction of adopted agents over time). (b,d) Numerical
  estimate of $g(\rho)$, obtained from $\rho(t)$ by inverting
  Eq.~(\ref{eq:PopGen}). Dashed curves correspond to the fit of Eq.~(\ref{eq:bassSol}) to
  $\rho(t)$.  Estimations of $G$ correspond to the area between the horizontal gray line ($g(\rho)=\hat{a}$) and the solid (${\tilde G}$) or dashed ($\hat{G}$) curves in (b,d).  Results: Bass $G=0.397,
  L=0.999,\hat{G}=0.415,\tilde{G}=0.400$; Threshold $G=0.347,L=0.988,  \hat{G}=0.314,\tilde{G}=0.352$.
}
\label{fig:Bass-thre-sf}
\end{figure*}

\begin{figure}[!t]
\includegraphics[width=0.49\columnwidth]{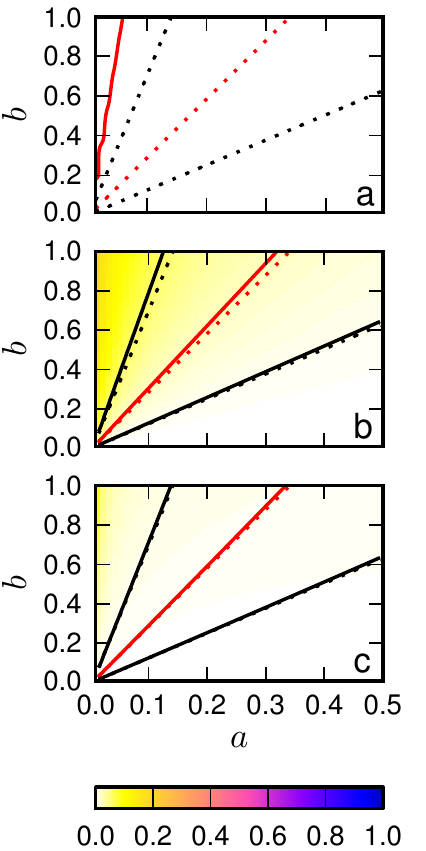} 
\includegraphics[width=0.49\columnwidth]{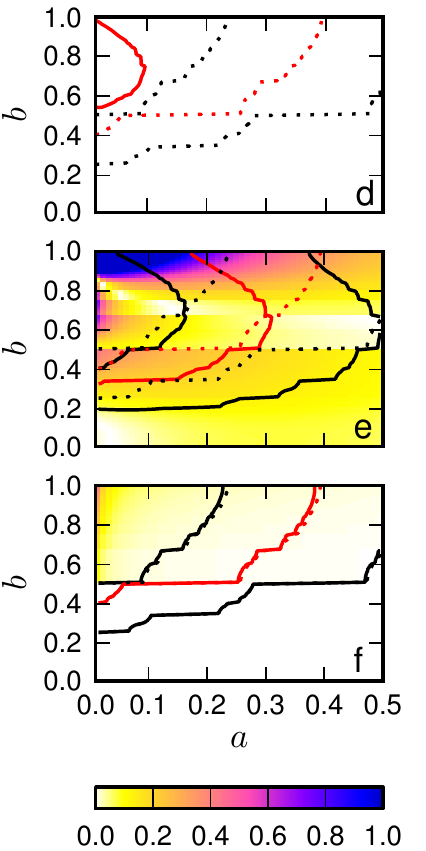} 
\caption{(color online) Strength of endogenous factors $G$ in the Bass [Eq.~(\ref{eq.bass}),
  panels a,b,c] and
  threshold [Eq.~(\ref{eq.threshold}), panels d,e,f]  models for  different parameters $a$ and $b$. The
  dashed lines correspond to values of $a,b$ for which $G=1/2$ (red), $G=1/3$ (black below
  red), and $G=2/3$ (black above red), computed from Eq. (\ref{eq.GAME}). The different panels show the estimations based on
  $L$ (a,d), $\hat{G}$ (b,e), and $\tilde{G}$ (c,f). Solid lines indicate values of $a,b$
  for which values $1/2, 1/3,$ and $2/3$ were obtained and should be compared to the corresponding dashed lines. The
  color code indicates the relative errors between the true value $G$ and the estimated values
  $\hat{G}$ (b,e) and $\tilde{G}$ (c,f). The model dynamics was simulated for scale-free networks with the same parameters as in Fig. \ref{fig:Bass-thre-sf}. 
}
\label{fig:Bass-thre-sf-error}
\end{figure}

\section{Time series estimators} \label{sec:methods}

In reality one usually has no access to information on individual agents and only the
aggregated curve $\rho(t)$ is available. This means that $G$ can not be estimated by Eqs. (\ref{eq:def}) or (\ref{eq.GAME}).
Here we propose and critically discuss the accuracy of three different methods to estimate $G$ from the S-curve $\rho(t)$ obtained from either empirical or surrogate data. All methods are inspired by the simple population model discussed
above, but can be expected to hold also in more general cases. Below we summarize the main idea of the three methods, details on the implementation appear in SM. III.


{\it Method 1, fit of S- and exponential curves:} We fit Eq.~(\ref{eq:bassSol}) by minimizing the Least-Square error with respect to the observed timeseries in the two limiting cases: (i) $a=0$, symmetric S-curve (endogenous factors only) and (ii) $b=0$,  exponential curve (exogenous factors only). 
Assuming normally distributed errors (which generically vary in time) we calculate the likelihood of the data given each model~\cite{Hastie2009}.
The normalized likelihood ratio $L$ of the two models indicates which curve provides a better description of the data~\cite{Burnham2002}.
The critical assumption in this method (to be tested below) is to consider the value of $L$ as an indication of the predominance of the corresponding factor, i.e $L>0.5$ indicates stronger exogenous factors $(G>0.5)$ and $L<0.5$ stronger endogenous factors $(G<0.5)$. 
This method does not allow for an estimation of $G$, but it provides an answer to the question of the most relevant factors.
The two simple one-parameter curves are unlikely to precisely describe many real adoption curves~$\rho(t)$. 
However, we expect that they will distinguish between cases showing a rather fast/abrupt start at $t_0$ (as in the exponential/exogenous case) from the ones showing a slow/smooth start (as in the S-curve/endogenous case). 
For this distinction, the $t\gtrapprox 0$ is the crucial part of the $\rho(t)$ curve because for $t \rightarrow \infty$ the symmetric S-curve approaches
$\rho=1$ also exponentially.


{\it Method 2, fit of generalized S-curve:} We fit Eq.~(\ref{eq:bassSol}) by
  minimizing the Least-Square error with respect to the timeseries and obtain the
  estimated parameters $\hat{a}$ and $\hat{b}$.
By inserting these parameters in Eq.~(\ref{eq:Gextintbass}) we compute $\hat{G}$ as an estimation of $G$.


{\it Method 3, estimation of $g(\rho)$:} We estimate $g(\rho)$ from Eq.~(\ref{eq:PopGen}) by calculating a (discrete) time derivative
$\dot{\rho}$ at every point $\rho(t)$. From a (smoothed) curve of $g(\rho)$ we consider
$g(0)$ to be the exogenous factors, write $g^{\text{endo}}=g(\rho)-g(0)$ and obtain an
estimation ${\tilde G}$ of $G$ from Eq.~(\ref{eq:G}). The advantage of this non-parametric method is that it is not a
priory attached to a specific $g(\rho)$ and therefore it is expected to work whenever
a population dynamics equation~(\ref{eq:PopGen}) provides a good approximation of the
data.

\begin{figure*}[!bt]
\includegraphics[angle=-90,width=2.\columnwidth]{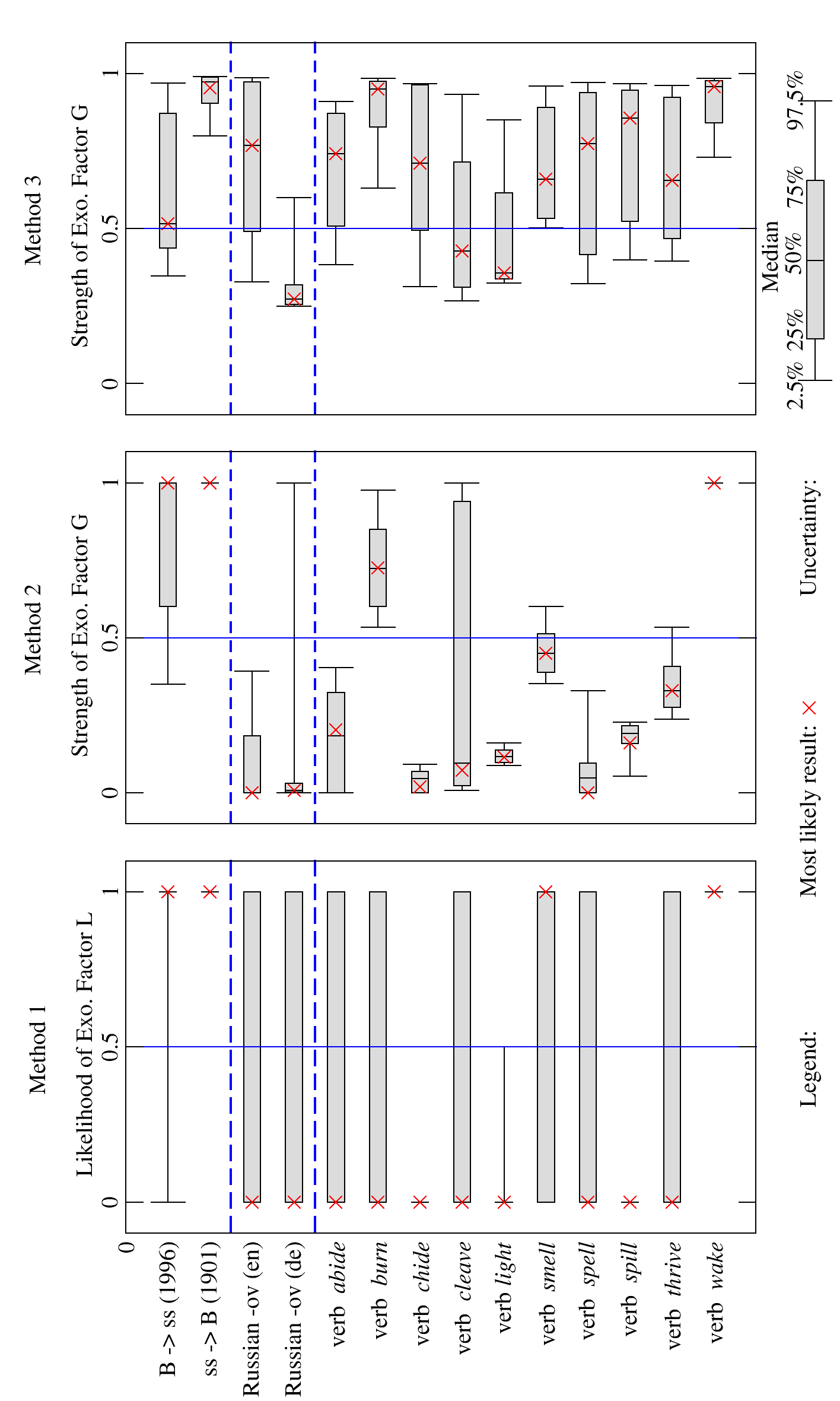}
\caption{(color online) Estimation of the strength of exogenous factors in empirical data.  The red X
  indicates the estimated value obtained using the complete database. The box-plots (gray
  box and black bars) were computed using bootstrapping and quantify the uncertainty of the
  estimated value (from left to right, the horizontal bars in the boxplot  indicate the
  $2.5\%,25\%,50\%,75\%,$ and $97.5\%$ percentile). Panels (a)-(c) show the estimations
  based on the three methods proposed in Sec.~\ref{sec:methods}.  (a) Method 1: the likelihood 
  ratio L of the exponential fit (exogenous factors) in relation to the symmetric S-curve
  fit (endogenous factors). (b) Method 2: estimation $\hat{G}$ based on the fit of
  Eq.~(\ref{eq:bassSol}) and on Eq.~(\ref{eq:Gextintbass}). Method 3: estimation $\tilde{G}$
  based on the general population dynamics model~\ref{eq:PopGen} (see SM. III for details
  on the implementation of the 3 methods and for figures of individual adoption curves).}
\label{fig:data}
\end{figure*}

\section{Application to network models} \label{sec:models}

Here we investigate time series $\rho(t)$ obtained from simulations of models in
which we have access to the microscopic dynamics of agents. Our goal is to measure $G$ on different models and to test
the estimators ($L,\tilde{G},\hat{G}$) defined in the previous
section. We consider two specific network models in the framework described in Sec.~\ref{ssec.networks},
which are defined fixing the network topology (in our case random scale-free) and the function $F_{k,m}$ (the adoption rate of an
agent having $m$ out of $k$ neighbours that already adopted) as~\cite{Gleeson:2013,Newman:2010}:

\begin{equation}\label{eq.bass}
\text{Bass model: }  F_{k,m} = a + b \frac{m}{k},
 \end{equation}

\begin{equation}\label{eq.threshold}
\text{Threshold: } F_{k,m} = \begin{cases} a, & m/k < 1-b\\ 1, & m/k \geq 1-b  \end{cases}.
\end{equation}

In both cases, when no infected neighbor is present ($m=0$), the rate is $F_{k,0}=a$
and therefore the parameter $a$ controls the strength of exogenous factors. Analogously, $b$
controls the increase of $F_{k,m}$ with $m$ and therefore the
strength of endogenous factors. Given a network and values of $a$ and $b$, we obtain
numerically both the timeseries $\rho(t)$  (using the AME formalism~\cite{Gleeson:2013,Gleeson:2011}, SM. IIC),  and the
strength of exogenous factors $G$ from Eq.~(\ref{eq.GAME}). Typically these models cannot be reduced to a one-dimensional population dynamics model and therefore the estimators $\hat{G}$ and $\tilde{G}$ (based on $\rho(t)$) differ
from the actual $G$. As a test of our methods, we compare the exact $G$ to $L$, $\hat{G}$ and ${\tilde G}$.

In Fig.~\ref{fig:Bass-thre-sf} we apply our time-series analysis to the two
models defined above with parameters $a=0.1, b=0.5$. Method 1 provides $L>0.5$ in both cases, incorrectly identifying that the exogenous factor is stronger. Furthermore, $\tilde{G}$ (Method 3)
provides a better estimation of $G$ than $\hat{G}$ (Method 2). This 
is expected since the estimation $\hat{G}$ is based on a straight line estimation of
$g(\rho)$ , $(\hat{a}+\hat{b}\rho)$, while $\tilde{G}$ admits more general function, see
Fig.~\ref{fig:Bass-thre-sf}, (b,d).
The estimations are better for the Bass model than for the threshold dynamics, consistent
with the better agreement between $\rho(t)$ and the fit of Eq.~(\ref{eq:bassSol}) in panel
(a) than in panel (c). 

In Fig.~\ref{fig:Bass-thre-sf-error} we repeat the analysis of Fig.~\ref{fig:Bass-thre-sf}
varying the parameters $a,b$ in Eqs.~(\ref{eq.bass}) and (\ref{eq.threshold}), while Eq. (\ref{eq.GAME}) gives the true value of $G$. The parameter space $a,b$ is divided in two regions: one for
which the exogenous factors dominate $G>0.5$ (below the red dashed line $G=0.5$) and one for which
the endogenous factors dominate $G<0.5$ (above the red dashed line $G=0.5$). In the Bass
dynamics the division between these regions  corresponds to a smooth (roughly straight)
line. In the threshold model a more intricate curve is obtained, with plateaus on rational
values of $b$ reflecting the discretization of the threshold dynamics in
Eq.~(\ref{eq.threshold})  (particularly strong for the large number of agents with few
neighbors). A strong indication of the limitations of the
$L$ and $\hat{G}$ estimators is that the $L=0.5$ (panel d) and $\hat{G}=0.5$ (panel e) lines show non-monotonic growth in the
$a,b$ space.  This artifact disappears using the ${\tilde G}$ estimator.   
Regarding the relative errors of the methods 2 and 3 (colour code), the results confirm
that $\tilde{G}$ is the best method and provides a surprisingly accurate estimation of $G$. Comparing the different models, the estimations for Bass are better than for the threshold dynamics (for the same parameters $(a,b)$). The
minimum errors are obtained for $b\approx0$  while for $a\approx0$ maximum errors for both methods are observed. 

%
%
%
%
%

\section{Application to data} \label{sec:data}
We now turn to the analysis of empirical data taken from the Google-ngram corpus~\cite{Michel:2011,Lin:2012}, see Ref.~\cite{Ghanbarnejad2014} and SM. I. We focus on the three cases reported in Fig.~\ref{fig.1}: 

{\it a. German orthographic reforms:} The 1996 orthography reform aimed to simplify the spelling of the German language based on
phonetic unification. According to this reform, after a short vocal one should write ``ss'' instead of ``\ss'', which predominated since the previous reform in 1901. This rule makes up over $90\%$ of the words changed by the reform \cite{wiki:GermanOrthography}. We combine all words affected by this rule to estimate the
strength of adoption of the orthographic reform, i.e., $\rho(t)$ is the fraction of
word tokens in the list of affected words written with ``ss''. Although following the reform was obligatory at schools, strong resistance against it led to debates even in the Federal Constitutional Court of Germany \cite{Johnson:2005}. For example, ``six years after the reform, $77\%$ of Germans consider the spelling reform not to be sensible \cite{wiki:GermanOrthography}''. These debates show that besides the exogenous pressure of language academies, endogenous factors can be important in this case also, either {\it for} or {\it against} the change.

{\it b. Russian names:} Since the $19$th century there have been different systems for the romanization of Russian names,
i.e. for mapping names from the Cyrillic to the Latin alphabet \cite{wiki:RomRuss}. These systems can be seen as exogenous factors. Alternatively, imitation from other authors can be considered as endogenous factors. All of the systems suggest a unique mapping from letter
``\foreignlanguage{russian}{в}'' to ``v'' (e.g., \foreignlanguage{russian}{Колмогоров} to Kolmogorov). Variants to
this official romanization system are ``ff'' or ``w''  (e.g., Kolmogorow and Kolmogoroff) which were used in different languages such as German and English. Here we study an ensemble
of $50$ Russian names ending in either
``-\foreignlanguage{russian}{ов}'' or ``-\foreignlanguage{russian}{eв}'' that were used often in English (en) and German (de). For each of these two languages, we combine all words (tokens) in order to obtain a single curve $\rho(t)$ measuring the adoption of the ``v'' convention.

{\it c. Regularization verbs in English:} A classical studied case of grammatical changes is regularization of English verbs
\cite{Lieberman:2007,Pinker:1999}. From 177 irregular
verbs in Old-English, 145 cases survived in Middle English and only 98 are still alive
\cite{Lieberman:2007}. Irregular verbs coexist with
their regular (past tense written by -ed) competitors, even if dictionaries may only present irregular forms \cite{Michel:2011}. Having an easier grammar rule or a rule aligned with a larger grammatical class are good motivations to use more often regular forms.  Other potential exogenous factors which favour works against regularization can be dictionaries and grammars. However, there are also cases of verbs that become irregular \cite{Michel:2011,Cuskley2014}. We analyse $10$
  verbs that exhibit the largest relative change. In $8$ cases regularization is observed.

\begin{figure*}[!t]
\includegraphics[width=2\columnwidth]{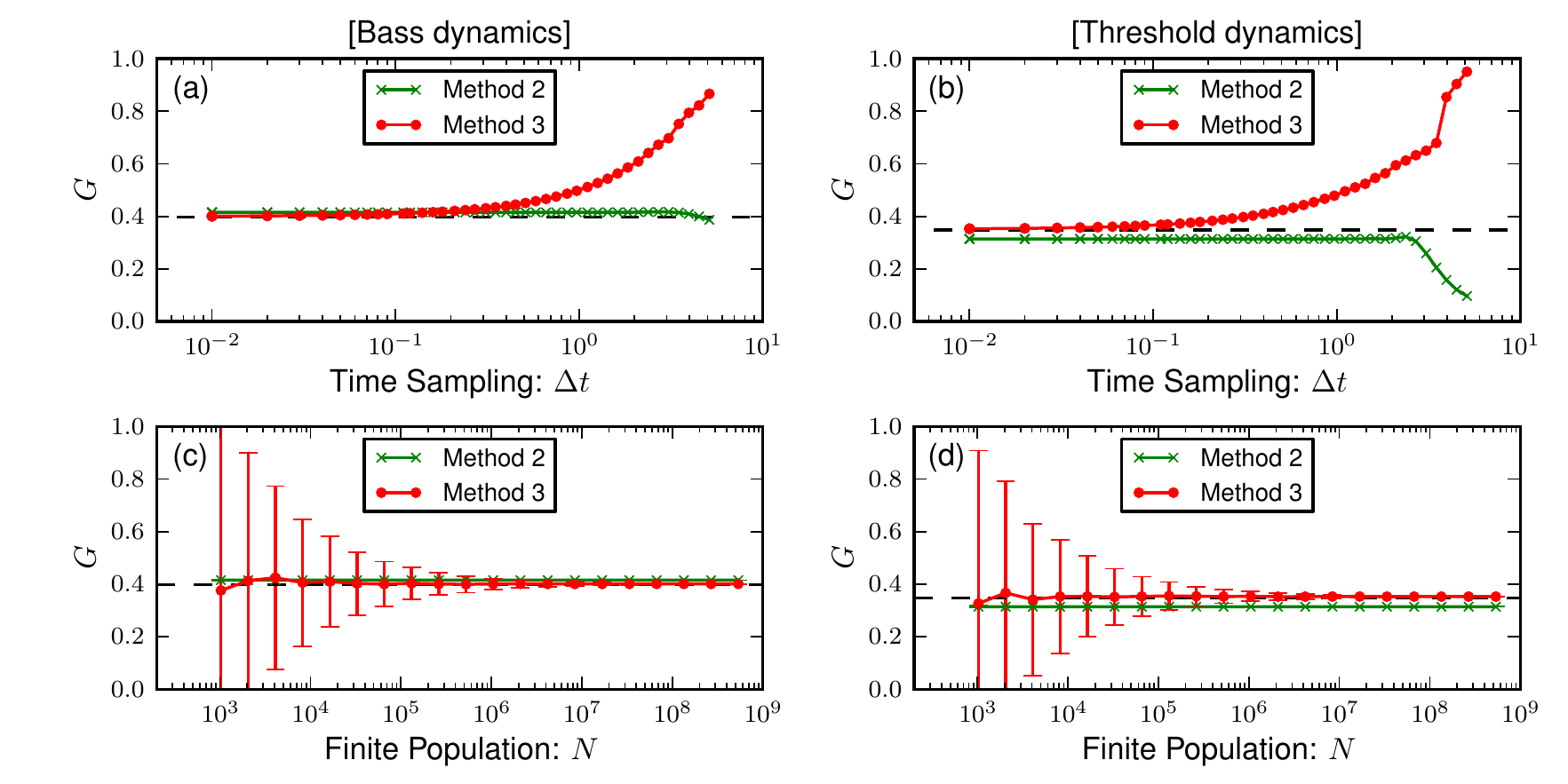} 
\caption{(color online) Method 2 is more robust against perturbations than Method 3.  Estimation of G in
    undersampled versions of the timeseries used in Fig. (2) for Bass (left)
 and threshold (right) dynamics.  The true  $G$ [Eq. (7)] is shown as a dashed line and
 Methods 2 and 3 are shown by symbols.
(a,b) Undersampling in time:  achieved by varying the time-resolution $\Delta t$ of the
timeseries, i.e., we sample $\rho(t)$ at times $\rho(t_0), \rho(t_0+\Delta t),
\rho(t_0+2\Delta t), \ldots$ .  Resolution increases for $\Delta t\rightarrow 0$.
(c,d) Undersampling of the population $N$. The surrogate time series $\rho(t)$ in
Fig.~\ref{fig:Bass-thre-sf} assume $N\rightarrow \infty$. We consider time series for
which only a finite population $N$ is 
observed. The observed fraction of adopters is determined from $N$ independent Bernoulli trials
with probability $\rho(t)$. This corresponds to adding noise to each data point
$\rho(t)$. Resolution increases for $N \rightarrow \infty$.  For each $N$, we plot the average and
standard deviation of $G$ computed over $1,000$ trials.} 
\label{fig:Bass-thre-resolution}
\end{figure*}

Besides the linguistic and historical interest in these three cases, there are also two
practical reasons for choosing these three simple spelling changes: (i) they
provide data with high resolution and frequency; and (ii) they allow for an unambiguous
identification of ``competing variants'', a difficult problem in language change~\cite{Hruschka:2009}. The last point allows us to concentrate on the
relative word frequency (as defined in the caption of Fig. \ref{fig.1}) which we identify with the relative number of adopters $\rho(t)$ in the
models of previous sections.  The advantage of investigating relative 
frequencies, instead of the absolute frequency of usage of one specific 
variation, is that they are not affected by absolute changes in the usage of the 
word.

Fig.~\ref{fig:data} shows estimations of the strength of exogenous factors G (using the methods of Sec.~\ref{sec:methods}) in the three examples of linguistic change described above. In line with the definition proposed in Sec.~\ref{sec:GenFram}, $G$ is interpreted as the fraction of adoptions because of exogenous factors. Besides the most-likely estimation obtained for the complete datasets
(red X), we have performed a careful statistical analysis (based on bootstrapping)  in order to determine the
confidence of our estimations (gray box plots). We first discuss the performance of the
three methods:

{\it Method 1:} The estimation of the likelihood $L$ that the exponential fit (exogenous factors)
  is better than the symmetric S-curve fit (endogenous factors)  resulted almost always in a
  categorical decision (i.e., $L=0$ or $L=1$). This is explained by the large amount of
  data that makes any small advantage for  one of the fits to be
  statistically significant. Naively, one could interpret this as a clear selection of the
  best model. However, our bootstrap analysis shows that in most cases the decision is not
  robust against small fluctuations in the data (gray boxes fill the interval $L\in [0,1]$). In these cases our conclusion is that the method is
  unable to determine the dominant factors (endogenous or exogenous).

{\it Method 2:} It generated the most tightly constrained estimates of G. The precision of the estimations of the
  strength of the  exogenous factors $G$ varied from case to case but remained typically
  much smaller than $1$ (with the exception of the
  verb {\it cleave}). In all cases for which Method 1 provided a definite result, Method 2 was consistent with it. This is not completely
  surprising considering that   the fit of the curve used in method 2 has as limiting
  cases the curves used in the fit  by Method 1. The advantage of Method 2 is that it
  works in additional cases (e.g., the Russian names), it provides an estimation of $G$
  (not only a decision whether $G>0.5$), and it distinguishes cases in
  which both factors contribute equally (verb {\it smell}) from those that data is unable
  to decide (verb {\it cleave}). 

{\it Method 3:} The results show large uncertainties and are shifted towards large values
of $G$ (in comparison to the two previous methods). In the few cases showing narrower
  uncertainties, an agreement with Method 2 is obtained in the estimated $G$ (verbs {\it wake}
  and {\it burn}) or in the tendency $G<0.5$ (Russian names in German). However, for most
  of the cases the uncertainty is too large to allow for any conclusion. The reason of
  this disappointing result is that Method 3 is very sensitive against fluctuations. For instance, it requires the computation of the temporal
  derivative of $\rho$. In simulations this can be done exactly and the method provided the best results in Sec.~\ref{sec:models}. However in empirical data,  discretization is unavoidable (in our case we have yearly resolution). Furthermore, fluctuations in the time-series become magnified when
  discrete time differences are computed (see SM. IIIC for a description of the
  careful combination of data selection and smoothing used in our data analysis). In
    order to test these hypotheses, in
  Fig.~\ref{fig:Bass-thre-resolution} we test the robustness of Methods 2 and 3 against
  discretization in time -- panels (a) and (b) -- and population -- panels (c) and (d) --
     for the  model systems treated in Sec.~\ref{sec:models}.  We observe that Method 3 is
     less robust than Method 2, showing a bias towards larger $G$ for temporal
     discretizations and broad fluctuations for population discretizations.
     These findings can be expected to hold for other types of noise and are consistent
     with our observations in the  data.

We now interpret the results of Fig.~\ref{fig:data} for our three examples (see
SM. Figs. (1-4) for the adoption curves of individual words):

{\it a.} Results for the {\bf German orthographic reform} indicate a stronger presence of
  exogenous factors, consistent with the interpretation of the (exogenous) role of language
  academies in language change being dominant.

{\it b.}  The {\bf romanization of Russian names} indicates a prevalence of endogenous
  factors. Most systems that aim at making the romanization uniform have
  been implemented when the process of change was already taking place (The change starts
  around $1900$ and first agreement is from $1950$). Moreover, the
  implementation of these international agreements is expected to be less efficient than the legally binding decisions of language academies (such as in orthographic reforms).

{\it c.}  The {\bf regularization of English verbs} show a much richer behavior. Besides some
  unresolved cases (e.g., the verb {\it cleave}) the general tendency is for a predominance of
  endogenous factors (e.g., the verbs {\it spill} and {\it light}), with some exceptions
  (e.g., the verb {\it wake}). 

%
%
%
%
%

\section{Discussions and Conclusions}

In summary, in this paper we combined data analysis and simple models to quantitatively investigate S-curves of vocabulary replacement. Our data analysis shows that linguistic changes do not follow universal S-curves (e.g., some curves are better described by an exponential than by a symmetric S-curve and fittings of Eq.~(\ref{eq:bassSol}) leads to different values of $\hat{a}$ and $\hat{b}$). These conclusions are independent of theoretical models and should be taken into account in future quantitative investigations of language change.

Non-universal features in S-curves suggest that information on the mechanism underlying the change can be obtained from these curves. To address this point, we considered simple mechanistic models of innovation adoption and three simplifying assumptions (identical agents, monotonic change, and constant strength of exogenous factors). We introduced a measure (Eq.~(\ref{eq:def})) of the strength of exogenous factors in the change and we discussed three methods to estimate it from S-curves.
Our results show a connection between the shape of the S-curves and the strength of
the factors (Fig. \ref{fig:Bass-thre-sf-error}). Exogenous factors typically break symmetries of the microscopic dynamics and lead to asymmetric S-curves. Thus the crucial point in all methods is to quantify how abrupt (exogenous) or smooth (endogenous) the curve is at the beginning of the change. We verified that both our proposed measure and methods correctly quantify the role of exogenous factors in binary state network models. In empirical data, the finite temporal resolution and other fluctuations have to be taken into account in order to ensure the results of the methods are reliable. These findings and the methods introduced in this paper -- data analysis and measure of exogenous factors -- can be directly applied also to other problems in which S-curves are observed~\cite{Rogers:2010,Vitanov:2012,Bass:1969,Bass:2004}.

S-curves provide only a very coarse-grained description of the spreading of linguistic
innovations in a population. For those interested in understanding the spreading
mechanism, the relevance of our work is to show that S-curves can be used to discriminate
between different mechanistic descriptions and to quantify the importance of different
factors known to act on language change. In view of the proliferation of competing models
and factors, it is essential to compare them to empirical studies, which are often limited
to aggregated data such as S-curves. Furthermore, quantitative descriptions of S-curves
quantify the speed of change and predict future developments. These features are
particularly important whenever one is interested in favoring convergence (e.g., the
agreement on scientific terms can be crucial for scientific progress~\cite{Knapp:2007} and
dissemination~\cite{Bentley:2012}).

\section*{Acknowledgements}

We thank J. C. Leit\~ao for the careful reading of the manuscript. 

\bibliography{paper}

\end{document}